\documentclass[%
twocolumn,
 reprint,
superscriptaddress,
 amsmath,amssymb,
 aps,
]{revtex4}

\usepackage{graphicx}

\usepackage{dcolumn}
\usepackage{bm}
\usepackage{tabularx}
\usepackage{amsmath}
 
\begin{document}

\preprint{APS/123-QED}

\title{Electric circuit model of microwave optomechanics}
\author{Xin Zhou}
\email{Corresponding Author: xin.zhou@iemn.fr}
\affiliation{CNRS, Univ. Lille, Centrale Lille, Univ. Polytechnique Hauts-de-France, IEMN UMR8520, Av. Henri Poincar\'e, Villeneuve d'Ascq 59650, France}
\author{Dylan Cattiaux}
\affiliation{Univ. Grenoble Alpes, Institut N\'eel - CNRS UPR2940, 25 rue des Martyrs, BP 166, 38042 Grenoble Cedex 9, France}
\author{Didier Theron}
\affiliation{CNRS, Univ. Lille, Centrale Lille, Univ. Polytechnique Hauts-de-France, IEMN UMR8520, Av. Henri Poincar\'e, Villeneuve d'Ascq 59650, France}
\author{Eddy Collin}
\affiliation{Univ. Grenoble Alpes, Institut N\'eel - CNRS UPR2940, 25 rue des Martyrs, BP 166, 38042 Grenoble Cedex 9, France}

\date{\today}

\begin{abstract}
We report on the generic {\it classical electric circuit modeling} that describes standard single-tone microwave optomechanics. 
Based on a parallel $RLC$ circuit in which a mechanical oscillator acts as a movable capacitor, derivations of analytical expressions are presented, 
including key features such as the back-action force, the input-output expressions, and the spectral densities associated, all in the classical regime.
These expressions coincide with the standard quantum treatment performed in optomechanics when the occupation number of both cavity and mechanical oscillator are large.
Besides, the derived analytics transposes optical elements and properties into electronics terms, which is mandatory for quantitative measurement and design purposes.
Finally, the direct comparison between the standard quantum treatment and the classical model addresses the bounds between quantum and classical regimes, highlighting the features which are truly quantum, and those which {\it are not}.

\end{abstract}

\maketitle


\section{\label{sec:level1}Introduction}

In the last decades, great scientific successes have been achieved in cavity optomechanics, which uses laser photons to explore the interaction between optical fields and mechanical motion \citep{aspelmeyer2014cavity}. 
Cavity optomechanics allows to cool down suspended micro-mirrors and to excite cold atom clouds through the radiation-pressure effect, which enables the investigation of mechanical systems in the quantum regime \citep{arcizet2006radiation, brennecke2008cavity, schliesser2009resolved}.
Optical forces also offer a method to enhance the resolution of nano-mechanical sensors through the optical spring effect \citep{yu2016cavity}.
Optomechancial platforms are thus both {\it model systems} containing rich physics to be explored \citep{aspelmeyer2014cavity}, but also {\it unique sensors} able to detect extremely tiny forces/displacements, especially in the quantum regime as foreseen in the 80's \citep{caves1980measurement}.
Besides, the amazing sensitivity of optomechanics for displacement detection lead recently to the {\it tour de force} detection of the long-thought gravitational waves \citep{abbott2016properties}.

Inspired by the achievements of circuit quantum electrodynamics (QED), researchers 
eventually started to use
microwave photons confined in a superconducting resonator to probe micro/nano-mechanical oscillators, leading to the new experimental field of {\it microwave optomechanics} \citep{regal2008measuring}. 
It inherits the abundant physical and technological properties emerging from cavity optomechanics, and benefits from the capabilities of microwave circuit designs. 
Especially, low temperature experiments performed in this wavelength range allow the use of quantum electronics components such as Josephson parametric amplifiers \citep{teufel2009nanomechanical}. 
Building on microwave optomechanical schemes, sideband cooling of mechanical motion down to the quantum ground state and entanglement of massive mechanical oscillators have been recently achieved 
\citep{teufelGround, ockeloen2018stabilized}. 
Moreover, microwave optomechanical platforms with cavity-enhanced sensitivity have been built, squeezing the classical thermal fluctuations of the mechanical element \citep{pernpeintner2014circuit, WeigCavityRT, huber2019squeezing}. The latter clearly demonstrates that optomechanics is not only reserved for frontiers experiments in quantum mechanics, but represents also a new resource for classical devices with novel applications.

Within circuit QED, Hamiltonian formulations adapted to various quantum circuits have been developed using the mathematical toolbox of quantum mechanics.
This has been achieved by quantizing (i.e. promoting to operators) variables of electrical engineering (e.g. voltage, current) and building the corresponding generic quantum circuit theory \citep{yurke1984quantum, vion2002manipulating, cottet2002implementation}. 
Signal propagation in coplanar waveguides (CPW) is thus described by propagating bosonic modes (i.e. photons, similarly to laser light), and $\textit{RLC}$ resonators by localized modes. Input-output theory and quantum noise formalisms \citep{gardiner2004quantum, gardiner2015quantum} are then required to describe driving fields and detected signals, in the framework of the quantum version of the fluctuation-dissipation theorem (applied here to superconducting circuits) \citep{goppl2008coplanar, clerk2010introduction}.
Among the properties of interest, quantum electric circuit theory enables to model non-linear features and dissipation, two crucial properties at the core of device performance and novel circuit development. For instance, quantum-limited Josephson amplifiers are described by modeling Josephson junctions as tunable nonlinear inductors \citep{vijay2009invited, zhou2014high}. 
The QED formalism has been recently detailed for a mechanical transducer \citep{zeuthen2018electrooptomechanical}, 
bridging quantum optics and quantum electronics; electro-opto-mechanical features are thus elegantly integrated within the framework of hybrid quantum systems. 

For microwave optomechanics, although its basic principle and relevant applications are not restricted to quantum mechanics, today the full classical circuit model has not yet been presented, despite 
useful pioneering discussions \citep{aspelmeyer2014cavity_book}. 
Therefore, the theoretical framework in use for all experiments, even when purely classical in essence, still relies on the Hamiltonian deduced from quantum optics \citep{aspelmeyer2014cavity}. 
This means both that the physical description is over-complexified, losing from sight the real (classical) nature of most properties; 
and also that the connection between circuit parameters and optomechanical properties is not clearly identified, while it is a need for design and optimization.

In this paper, we present the generic classical electric circuit modeling of microwave optomechanics. 
First in part \ref{sec:circuit}, we review the different $\textit{RLC}$ equivalent circuits at stake and then describe the framework of our formalism in section \ref{sec:soluce}.
In the following part \ref{sec:cavityForce}, we derive the cavity back-action force on the mechanics with its two components: the {\it dynamic} one (proportional to motion amplitude) that modifies the mechanical susceptibility, and the {\it stochastic} one (from the cavity current noise) that limits the mechanical mode fluctuations.
In part \ref{sec:output}, we then work out the corresponding input-output expressions with classical spectral densities.
In the last part \ref{sec:discussion} of the paper, we compare the classical description to the conventional quantum Hamiltonian method and discuss the
key ingredients of the models: namely {\it sideband asymmetry, the Heisenberg uncertainty}, and {\it the motion optimal detection limit} for which we define a (relative) {\it standard classical limit} (SCL), in analogy with the (ultimate) {\it standard quantum limit} (SQL). The major merit of this electric $\textit{RLC}$ circuit model lies in the fact that it gives 
a straightforward connection between experimentally relevant quantities and electric elements of the devices. 
It also offers a clear understanding of most features, demonstrating their classical nature (with no need to invoke quantum mechanics).

\section{\label{sec:model}Modeling of the classical electric analogue}

\subsection{\label{sec:circuit}Generic microwave optomechanical circuit}

\begin{figure}
\includegraphics[width=0.45\textwidth]{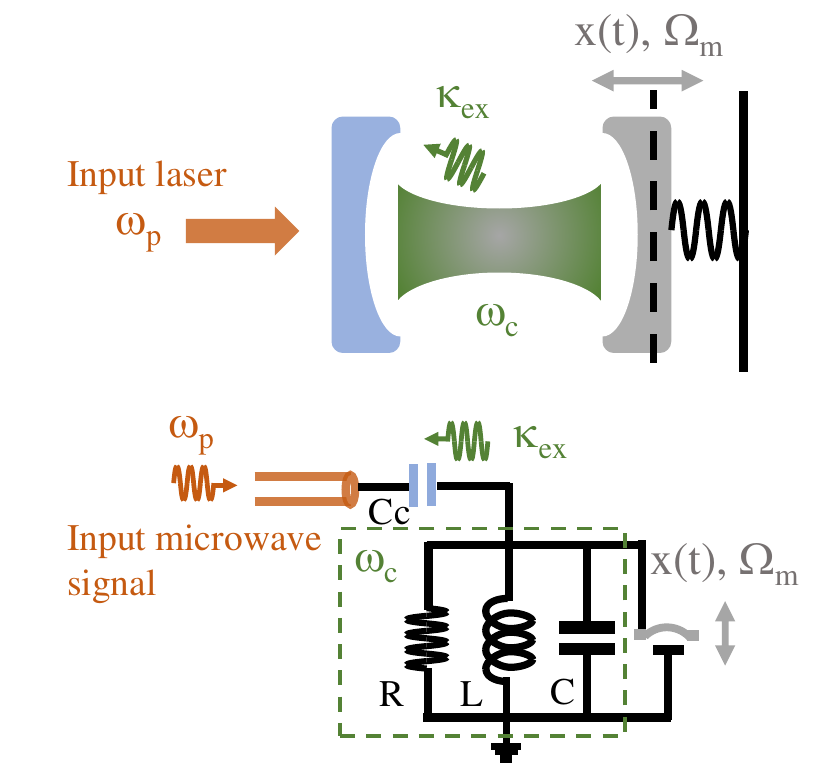}%
\caption{\label{fig:compare} Schematic diagram comparing a generic optomechanical system (top) and an electrical $\textit{RLC}$ circuit (bottom). See text for the color code. }
\end{figure}

In the electromechanical version of optomechanics, the mechanical element of the circuit is a {\it movable capacitor} which is part of an $\textit{RLC}$ resonator.
A schematic comparison is shown in Fig. \ref{fig:compare} in the reflection mode. 
The input laser is replaced by the microwave signal (orange), entering the optical cavity (green) through a semi-reflecting mirror for the former (blue), and the $\textit{RLC}$ resonator (green) through a capacitor $C_c$ for the latter (blue).
The movable element (gray) modulates the resonance frequency $\omega_c$ of the optical/microwave mode considered at a frequency $\Omega_m$.   
Part of the confined energy eventually leaks out at a rate $\kappa_{ex}$, back to the input (green sinusoidal arrow).

In practice, this resonant circuit can be realized in many ways: a quarter-wave coplanar waveguide (CPW) element \citep{regal2008measuring}, a microfabricated superconducting inductor-capacitor meander \citep{teufel2009nanomechanical}, or even a parallel plate capacitor shunted by a spiral inductor \citep{teufelGround}. 
{\it Any} resonator can be described by equivalent lumped $\textit{RLC}$ elements near a resonance; see e.g. Ref. \citep{goppl2008coplanar} for the case of CPW resonators, providing analytic expressions. For complex geometries however, finite element analysis is required (like e.g. in Ref. \citep{zhou2019chip}). 
In most experiments, the coupling to the outside is capacitive \citep{regal2008measuring,teufel2009nanomechanical,teufelGround,goppl2008coplanar,zhou2019chip}:
the resonator is almost isolated from the outside world, while the electromagnetic field from input/output waveguides is allowed to ``leak in-and-out'' through weakly coupled ports. This defines the microwave cavity element, in which the motion $x(t)$ of the mechanical object modulates the effective $C$.

\begin{figure}
\includegraphics[width=0.47\textwidth]{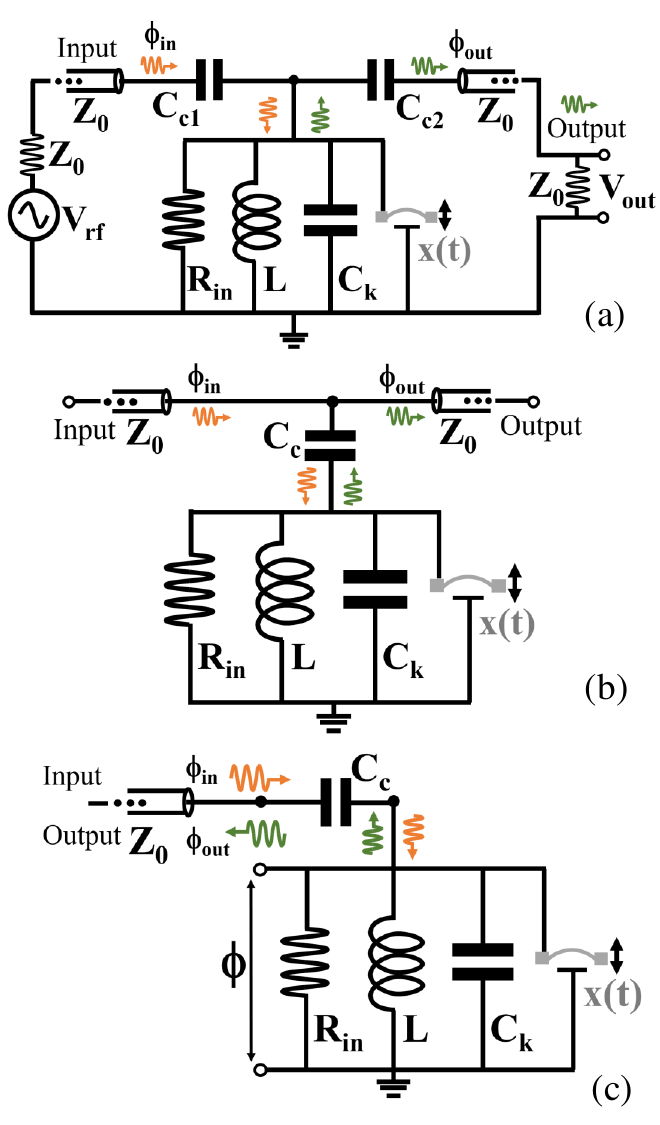}%
\caption{\label{fig:elementCircuit} The three equivalent electric circuits for microwave optomechanics with different input and output designs. In each, a mechanical element $x(t)$ capacitively couples to a parallel $\textit{RLC}$ circuit. (a) two-port, (b) bi-directional and (c) single-port (reflection design, a specific component is then required to separate incoming ant outcoming waves). The defined electric elements are discussed in the text, and the outside world, as seen from the circuit, is depicted only in (a). 
}
\end{figure}

In Fig. \ref{fig:elementCircuit} we thus show the three standard microwave setups. In (a) a two-port scheme, in which a 
lumped %
$\textit{RLC}$ parallel circuit couples to distinct input and output ports through different capacitors $C_{c1}$ and $C_{c2}$, yielding effective coupling rates $\kappa_1$ and $\kappa_2$ respectively;  %
they quantify how energy decays over time from the resonator to each port  \citep{hertzberg2010back, lei2016quantum}. The total coupling rate to the outside is thus defined as $\kappa_{ex}=\kappa_1+\kappa_2$. The internal damping rate of the circuit is modeled through $R_{in}$, leading to a $\kappa_{in}$ decay rate  %
(measuring the decay toward internal degrees of freedom). The total decay rate of the microwave mode is then $\kappa_{t}=\kappa_{ex}+\kappa_{in}$. %
The ports are realized by coaxial cables of characteristic impedance $\text{Z}_0$ connected to adapted elements, i.e. the outside impedance {\it seen from the port} is also $Z_0$ [see Fig. \ref{fig:elementCircuit} (a) input and output]: the voltage source %
$V_{\text{rf}} (t)= V_p \cos (\omega_p t)$ on the left has an output load of $Z_0$, while the detection of the $V_{\text{out}}$ voltage on the right is realized by an amplifier of input load $Z_0$ as well. $\omega_p$ is the (angular) microwave drive frequency, while $V_p$ is the applied amplitude. In Fig. \ref{fig:elementCircuit} (b) we show the electric schematic of a bi-directional coupling: the $\textit{RLC}$ resonator couples evanescently to a nearby transmission line with an effective capacitance $Cc$ \citep{palomaki2013coherent, teufelGround, zhou2019chip}.
This is strictly equivalent to scheme (a) when imposing $C_{c1}=C_{c2}=Cc/2$ (and thus $\kappa_{2}=\kappa_1=\kappa_{ex}/2$).
At last, the third scheme is shown in Fig. \ref{fig:elementCircuit} (c). Only one port is connected to the device, requiring thus the use of a specific nonreciprocal component (e.g. like a circulator) to separate the drive signal from the response (reflection mode) \citep{SillanpaaAmpification}. This is again equivalent to scheme (a) with $C_{c1}=C_{c}$ and no $C_{c2}$ (i.e. $\kappa_{2}=0$, $\kappa_{ex}=\kappa_1$). The problem at hand is solved below in terms of generalized fluxes $\phi(t)$ = $\int^t V(t') dt' $ \citep{clerk2010introduction}; the dynamics equation will be written for the $\phi$ corresponding to the $\textit{RLC}$ node, see Fig. \ref{fig:elementCircuit} (c). 
Incoming and outcoming traveling waves (equivalent to the laser signals in conventional optomechanics) are thus defined as $\phi_{in}$ (the pump tone, orange) and $\phi_{out}$ (the response, green) in Fig. \ref{fig:elementCircuit}. 

\subsection{\label{sec:soluce}Dynamics equation and solution}

We shall consider in the following a single port configuration [Fig. \ref{fig:elementCircuit} (c)], the extension to the other models being straightforward from what has been said above. Whenever necessary, this correspondence will be explicitly discussed.

The circuits shown in Fig. \ref{fig:elementCircuit} are a combination of transmission lines (the coaxial cables) and lumped elements ($RLC$, $Z_0$ impedances, and source).
The first step of the modeling is thus to get rid of the coaxial elements, in order to model an {\it ideal lumped circuit}. To start with, we consider the source $V_{\text{rf}}$ which generates the incoming wave $\phi_{in}$. 
In schemes (a) and (c) of Fig. \ref{fig:elementCircuit}, the drive port is terminated by an (almost) open circuit since the coupling capacitance is very small ($C_c \omega \, Z_0 \ll 1$). The incoming wave is thus almost fully reflected, and the standing wave voltage on the input capacitor is $V_d \approx 2 V_{\text{rf}}$ \citep{pozar2009microwave}. On the other hand for scheme (b), the transmission line is almost unperturbed by the coupling element $C_c$, and the incoming wave travels toward the output port (almost) preserving its magnitude; on the coupling capacitor we have $V_d \approx V_{\text{rf}}$.

\begin{figure}
\includegraphics[width=0.47\textwidth]{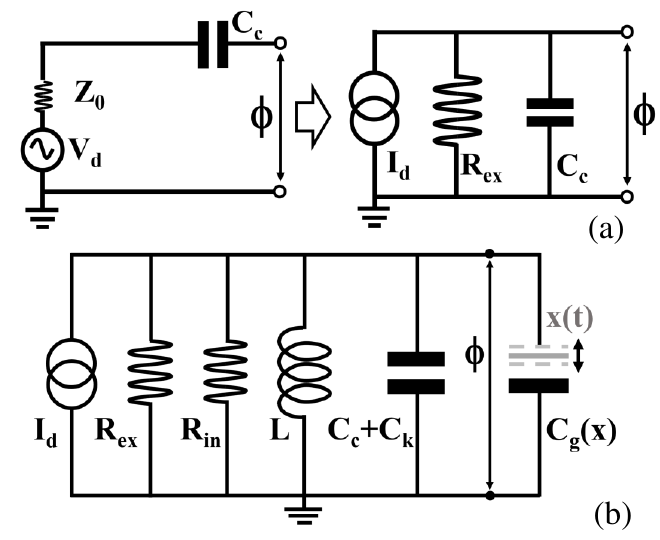}%
\caption{\label{fig:lcr_parelle} (a) Circuit schematics converting the series voltage source $V_{d}$ (impedance $Z_0$) that loads the capacitor $C_c$ into a parallel current source $I_{d}$; the equivalent resistance $R_{ex}$ is defined from $Z_0$ and $C_c$ (see text). (b) Equivalent parallel $RLC$ circuit with current source for the one-port optomechanical circuit [replacing Fig. \ref{fig:elementCircuit} (c)].
The mechanical contribution to the capacitance is explicitly defined as $C_g(x)$. }
\end{figure}

Applying Norton's theorem, we transform the series voltage source input circuit into a parallel $\textit{RC}$, which drives a total current $I_{d}$ across it. This is shown in Fig. \ref{fig:lcr_parelle} (a), with finally the total loaded $\textit{RLC}$ resonator in Fig. \ref{fig:lcr_parelle} (b). 
The effective components of the Norton drive circuit are defined from the real and imaginary parts of the complex admittance $Y_c(\omega) = [ Z_0 + 1 /(i C_c \omega) ]^{-1}$, in the limit $C_c \omega \, Z_0 \ll 1$ (weak coupling): 
\begin{eqnarray}
\label{eqn:element1}
\Re [Y_c(\omega)] &=& \frac{1}{R_{ex}} \approx (\omega_c C_c)^2\,Z_0 , \\
\Im [Y_c(\omega)] &\approx& i \omega_c C_c ,
\end{eqnarray}
with the approximation $\omega \approx \omega_c$ ($i$ is the {\it imaginary} unit). The current $I_{d}$ flowing into the resonator then writes:
\begin{equation}
\label{eqn:element2}
I_{d} \approx i \omega_c C_c \, V_{d}.
\end{equation}

The detected voltage is calculated from the current flowing through the amplifier's impedance $Z_0$. For circuits Fig. \ref{fig:elementCircuit} (a) and (c), this simply leads to: 
\begin{equation}
\label{eqn:element3}
V_{\text{out}} \approx - \omega_c^2 C_c Z_0 \, \phi,
\end{equation}
assuming again $\omega \approx \omega_c$. For circuit (b), the evanescent coupling leads to a loading composed of two impedances $Z_0$ in parallel (half of the signal is fed back to the voltage source):
\begin{equation}
\label{eqn:element3bis}
V_{\text{out}} \approx - \omega_c^2 C_c \frac{Z_0}{2} \, \phi.
\end{equation}

From our definitions of $\kappa_{ex}$, 
the subtlety of these different writings shall obviously be accounted for in our final expressions (see discussion of Section \ref{sec:output}).

The classical dynamics equation that describes this problem writes:
\begin{eqnarray}
\label{eqn:motion}
 &  & \!\!\!\!\!\!\!\!  \frac{d}{dt} \left( \left(C_{c}+C_{k}+C_{g}[x] \right) \frac{d \phi}{d t}\right)   \nonumber \\
 & +  & \left( \frac{1}{R_{ex}}+\frac{1}{R_{in}} \right) \frac{d \phi}{dt} + \frac{1}{L}\phi \nonumber \\
 & = & I_{d} + I_{\text{noise}} .
\end{eqnarray}
 
In Eq. (\ref{eqn:motion}), the fraction of the capacitance modulated by the mechanics is defined as $C_{g}(x)$ [see Fig. \ref{fig:lcr_parelle} (b)]. The Johnson-Nyquist  electric noise (at temperature $T_c$) seen by the cavity is modeled as a source $I_{\text{noise}}$, in parallel with the imposed drive circuit
generating $I_{d}$ \citep{clerk2010introduction}.

In the following, we will consider small motion. We therefore write $C_{g}(x) \approx C_{g}(0) +x(t) \times dC_{g}(0)/dx $, defining the total (static) capacitance $C_t=C_c+C_k+C_{g}(0)$ \citep{sulkko2010strong, sillanpaa2009accessing}. 
$C_{g}(0)$ corresponds to the contribution of the mobile element when at rest, while $C_c$ comes from the slight ``leakage'' of the cavity mode into the coaxial lines. The rates are then defined from the electronic components:
\begin{eqnarray}
\kappa_{ex} &=& \frac{1}{R_{ex} C_t} ,  \label{eqn:kappaex} \\
\kappa_{in} &=& \frac{1}{R_{in} C_t} .
\end{eqnarray}
In the two-port case, one simply defines $R_{1}$ and $R_{2}$ leading to $1/R_{ex}=1/R_{1}+1/R_{2}$, and similarly
$1/R_t=1/R_{ex} + 1/R_{in}$; we write the corresponding quality factors $Q_i = \omega_c/\kappa_i$ (with $i=in,ex,t$). Besides, the microwave resonance of the loaded $RLC$ circuit is given by $\omega_c = 1/\sqrt{L C_t}$.

Introducing the {\it coupling strength} $G= - d \omega_c / d x = - d \omega_c/d C_g \times d C_g/dx$, Eq. (\ref{eqn:motion}) can be re-written in the more compact form:
 
\begin{eqnarray}
\label{eqn:motionx}
 &  & \!\!\!\!\!\!\!\!  \left(1+\frac{2 G}{\omega_c} {x} \right) \frac{d^2  \phi }{dt^2}   \nonumber \\
 & +  & \left({\kappa_t}+\frac{2 G}{\omega_c}\frac{dx}{dt}\right)  \frac{d \phi}{dt} + {\omega_c}^2 \phi   \nonumber \\
 & = & \frac{I_{d}}{C_{t}}+ \frac{I_{\text{noise}} }{C_{t}},
\end{eqnarray}

\begin{figure}
\includegraphics[width=0.45\textwidth]{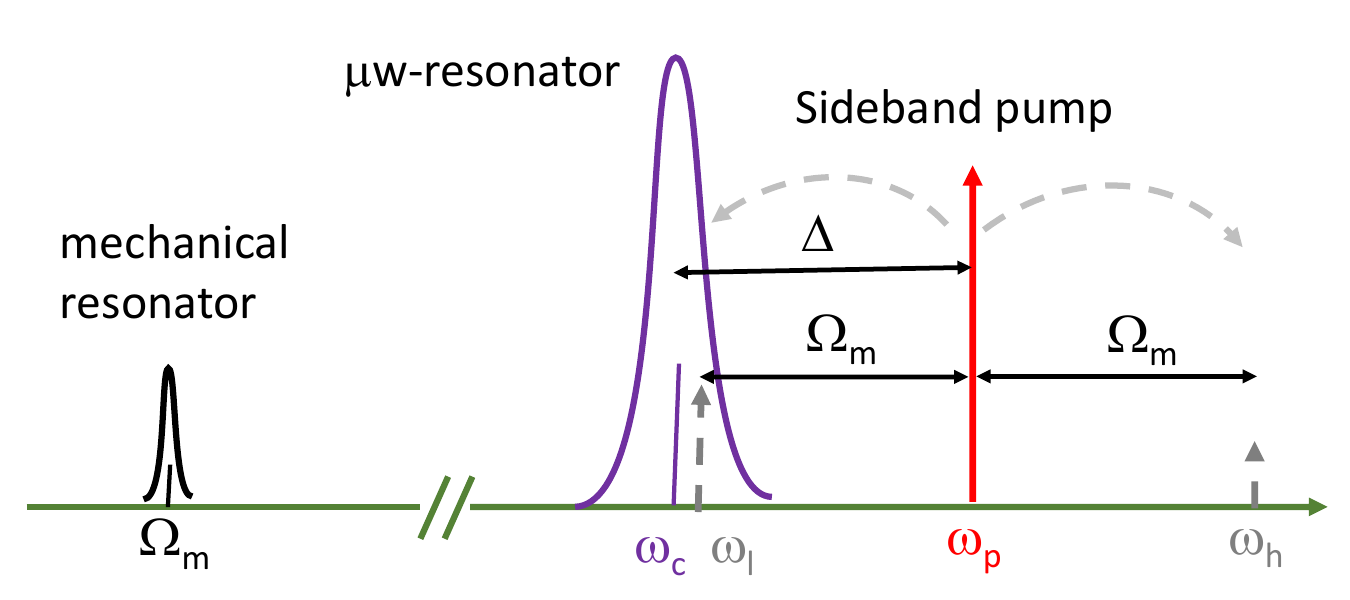}%
\caption{\label{fig:schematicFreq} Spectrum landscape: here, schematic example of ``blue sideband'' pumping (pump tone in red, $\Delta$ around $+\Omega_m$), in the resolved sideband limit ($\Omega_m \gg \kappa_t$, with $\kappa_t$ the width of the cavity response, here in violet). On the left, the Brownian mechanical peak around $\Omega_m$. In dashed gray, the mixing mechanism and the two main response peaks (around $\omega_l$ and $\omega_h$, see text).}
\end{figure}

The drive current writes $I_{d}(t)=\frac{1}{2} I_{p} e^{-i\omega_{p} t}+ \text{c.c.}$ with $\omega_{p}$ the frequency at which the microwave pumping is applied and $I_{p}$ its (complex) amplitude. 
We introduce the frequency detuning $\Delta = \omega_p-\omega_c$. 
From Eq. (\ref{eqn:element2}), $I_{p}$ is derived from the input voltage drive amplitude $V_{p}$ \cite{note}. The mechanical displacement is written as $x(t)=\frac{1}{2} x_0(t)   e^{-i \Omega_m t}+ \text{c.c.}$ with $x_0(t)$ the (complex) motion amplitude translated in frequency around $\Omega_m$, the mechanical resonance frequency of the movable element. This amplitude is a stochastic variable: the {\it Brownian motion} of the moving element thermalized at temperature $T_m$(see Fig. \ref{fig:schematicFreq}), in the absence of the back-action from the circuit. 

The terms where motion $x(t)$ multiplies flux $\phi(t)$ in Eq. (\ref{eqn:motionx}) then generate harmonics at $\omega_n = \omega_p + n \, \Omega_m$, with $n \in \mathbb{Z}$: this phenomenon is known as {\it nonlinear mixing}. We can thus find an exact solution using the {\it ansatz}:
\begin{equation}
\label{eqn:ansatz}
\phi (t) = \!\!\! \sum_{n=-\infty}^{+\infty} \!\!\! \phi_n (t)= \!\!\! \sum_{n=-\infty}^{+\infty} \!\! \frac{1}{2}\mu_n (t) e^{-i \left(\omega_p+ n \Omega_m\right) t}+ \text{c.c.}
\end{equation}
which, when injected in Eq. (\ref{eqn:motionx}) generates a system of coupled equations for the $\mu_n$ (complex) amplitudes.
In order to match the decomposition, the white noise component is thus naturally written as $I_{\text{noise}} = \sum_n \frac{1}{2} \delta I_n (t) e^{-i \left(\omega_p+ n \Omega_m\right) t}+ \text{c.c.} $ with $\delta I_n (t) $ the (complex) amplitude translated at frenquency $\omega_p+ n \, \Omega_m$. 

In practice, we are interested only in schemes where $\Delta \approx 0, +\Omega_m$ and $-\Omega_m$. 
The resonant feature brought in by the $RLC$ element implies that only the spectral terms the closest to $\omega_c$ in Eq. (\ref{eqn:ansatz}) will be relevant. 
In the so-called {\it resolved-sideband limit} where $\kappa_t \ll \Omega_m$, the amplitudes of these components fall off very quickly, and only the very first ones are needed to describe the systems' combined dynamics. Note however that the mathematical treatment performed here does not rely on this hypothesis, and it can be extended to compute the comb terms up to an arbitrary order in the non-resolved situation.

From the full comb, we thus keep only $n =-1, 0, +1$ which we rename in 'l' (lower), 'p' (pump) and 'h' (higher) respectively for clarity. Considering standard
experimental parameters in microwave optomechanics $\Omega_m \ll \omega_c$ \citep{SillanpaaAmpification, WeigCavityRT, teufelGround,zhou2019chip}, we also justify the assumption $\omega \approx \omega_c$ used in Eqs. (\ref{eqn:element1} - \ref{eqn:element3bis}). 

Eq. (\ref{eqn:motionx}) is solved by mimicking a {\it rotating wave approximation}, a method initially developed for resonances in atom optics and nuclear magnetic resonance \citep{rwa_paper}.
For our classical treatment, it simply means that we are concerned by the dynamics of each component of Eq. (\ref{eqn:ansatz}) only around its main frequency $\omega_n$ (what is called a ``rotating frame'', as opposed to the ``laboratory frame'' for the full signal), assuming time-variations of $x$ and $\mu_n$ to be slow (valid for high-$Q$ microwave and mechanical resonances). 
As stated above $\Omega_m \ll \omega_c$, which allows us to make the approximation $\omega_{l} \approx \omega_{p} \approx \omega_{h} \approx \omega_c$. Then, the flux amplitudes inside the cavity acting on the mechanical oscillator are:
\begin{align}
\label{eqn:ampInside}
\begin{split}
&\mu _p  (t)= +\frac{i}{2}\left(\frac{I_p}{\omega_c C_t}+\frac{\delta I_{p} (t) }{\omega_c C_t}\right) \chi_{p} ,
\\
&\mu _{l}(t)=+\frac{i}{2} \left(G \, x_0^{*}(t) {\mu _p}(t) +\frac{\delta I_{l}(t)}{\omega _{c} C_{t}} \right)\chi_{l} ,
\\
&\mu _{h}(t)=+\frac{i}{2} \left(G \, x_0(t) {\mu _p}(t) +\frac{\delta I_{h} (t) }{\omega _{c} C_{t}} \right)\chi_{h} ,
\end{split}
\end{align}
having defined:
\begin{align}
&\chi _{p}=\frac{1}{-i\Delta + \frac{\kappa_t}{2}} , \label{eqn:chiopt1}
\\
&\chi _{l}=\frac{1}{-i\left(\Delta-\Omega _m\right)+ \frac{\kappa_t}{2}} ,
\\
&\chi _{h}=\frac{1}{-i\left(\Delta+\Omega _m\right) + \frac{\kappa_t}{2}} , \label{eqn:chiopt3}
\end{align}
the {\it cavity susceptibilities} associated to each spectral component. 

When $\Delta \approx 0$, the pump tone 'p' is resonant with the cavity; the scheme is symmetric and the two 'l' and 'h' amplitudes are equivalent. We shall call 
this the ``green'' pumping scheme in the following.
When $\Delta \approx -\Omega_m$, the 'h' component is resonant with the cavity and the 'l' one is greatly suppressed. This is known as the ``red sideband'' pumping scheme. When $\Delta \approx +\Omega_m$, the situation is reversed and 'l' is resonant with the cavity and 'h' suppressed. This is the ``blue'' scheme schematized in Fig. \ref{fig:schematicFreq}. The meaning of the naming colors will be clarified in Section \ref{sec:output}.

\begin{figure}
\includegraphics[width=0.47\textwidth]{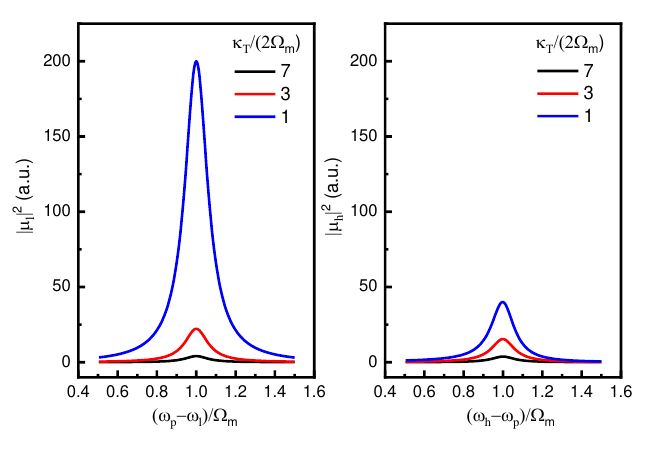}%
\caption{\label{fig:sideband} Schematic frequency dependence of the ${\mu}_{l}$ (left) and ${\mu}_{h}$ (right) components, under ``blue sideband'' pumping. Each curve corresponds to a different ratio of ${\kappa_t}/2\Omega_m$ (see legend: {\it not} in the sideband resolved limit), in the small $\mu_p$ drive limit (see text).}   
\end{figure}

The two satellite signals $\mu_l$ and $\mu_h$ at $\omega_p\pm\Omega_m$ generated by the pump tone $\mu_p$ are schematized in Fig. \ref{fig:sideband} for an arbitrary Brownian noise $x$, in the small drive limit. They correspond to energy up-converted from the pump 'p' (for 'h'), or down-converted (for 'l'). 
When the drive $\mu_p$ (i.e. $I_p$ or $V_{p}$) becomes large enough, {\it back-action} of the cavity onto the mechanical element has to be taken into account. This is derived in the next Section.

\subsection{\label{sec:cavityForce}Classical back-action of cavity onto mechanics}

The voltage bias on the mechanical element is responsible for a force $F_{\text{ba}}$, defined as the gradient of the electromagnetic energy:
\begin{equation}
F_{\text{ba}}(t) = +\frac{d}{dx} \left[\frac{1}{2} C_g(x)  \left( \frac{d \phi (t)}{dt}\right)^2 \right],
\end{equation}
using our notations (see Fig. \ref{fig:lcr_parelle}). This is the so-called {\it back-action} of the cavity onto the mechanical element.
The sign definition corresponds to the fixed electrode acting upon the mobile part. We are interested only in the component of this force that can drive the mechanics; we therefore define $F_{\text{ba}}(t) = \frac{1}{2} F_0(t)   e^{-i \Omega_m t}+ \text{c.c.}$ with $F_0(t)$ the (complex) force amplitude acting in the frame rotating at the mechanical frequency $\Omega_m$.
Re-writing $\phi$ in terms of the defined flux amplitudes, and keeping only the lowest order for the spatial derivative (small motion limit), we get: 
\begin{equation}
\label{eqn:baforce}
F_{0}(t) \approx + C_t \omega_c G \left[ \mu_p (t) \mu_l^{*} (t) +\mu_p^{*} (t) \mu_h (t)  \right].
\end{equation}  

From Eqs. (\ref{eqn:ampInside}), we immediately see that $F_0(t)$ will depend on both the motion amplitude $x_0(t)$ and the current noise components of the cavity $\delta I_n(t)$. Eq. (\ref{eqn:baforce}) is thus recasted in:
\begin{eqnarray}
\label{eqn:baforce2}
F_{0}(t) &\approx & + i \frac{G^2 }{\omega_c } \left( \frac{C_t \omega_c^2 \left|\mu_p \right|^2 }{2} \right)  \, \Big[ \chi_h - \chi_l^{*} \Big] \, x_0(t)  \nonumber \\
&+ &\!\!\!  i \frac{G}{2} \Big[ \delta I_{h}(t) \, \mu_p^{*} \, \chi_h - \delta I_{l}^{*}(t) \, \mu_p \, \chi_l^{*} \Big] ,
\end{eqnarray} 
with the first term the so-called {\it dynamic} component (proportional to $x_0$), and the last term the {\it stochastic component that is fed back} from the cavity onto the mechanical degree of freedom. In Eq. (\ref{eqn:baforce2}), the term $\delta I_p$ has been dropped at lowest order and $\mu_p \approx \frac{i}{2} \frac{I_p}{\omega_c C_t} \chi_p$ is now time-independent; only the noise current at the two sidebands is relevant. \\

The governing equation for the mechanical motion is expressed in the rotating frame as (neglecting again the fast dynamics): 
\begin{equation}
\label{eqn:xmotion}
-2 i \, \dot{x}_0(t)-i \Gamma_m x_0(t) = \frac{L_{0}(t)+F_{0}(t) }{m \Omega_m},
\end{equation}
with $\Gamma_m$ the mechanical damping rate, and $Q_m=\Omega_m/\Gamma_m$ the mechanical quality factor. $m$ is the mass of the moving element.
We write $L(t)$ the Langevin force (at temperature $T_m$), with $L_0(t)$ the component acting in the frame rotating at $\Omega_m$.
Injecting Eq. (\ref{eqn:baforce2}) into Eq. (\ref{eqn:xmotion}) and taking the Fourier transform, the solution can be written in the simple usual form:
\begin{equation}
\label{eqn:linear}
x_0(\omega) = \chi_m(\omega) \left[L_{0}(\omega) + \delta F_{0}(\omega) \right],
\end{equation}
where we have defined:
\begin{eqnarray}
\chi_m(\omega) & = & \frac{1}{2 m \Omega_m \left(-  \omega -i \frac{\Gamma_m}{2} \right) + \Sigma} , \label{eqn:chim} \\
\Sigma & = & -i \frac{G^2}{\omega_c} \left( \frac{C_t \omega_c^2 \left|\mu_p \right|^2 }{2} \right)  \, \Big[ \chi_h - \chi_l^{*} \Big] , \label{eqn:selfE}\\
\delta F_{0}(\omega)  & = & +i\frac{G}{2} \Big[ \delta I_{h}(\omega) \, \mu_p^{*} \, \chi_h - \delta I_{l}^{*}(\omega) \, \mu_p \, \chi_l^{*} \Big]. \label{eqn:noise}
\end{eqnarray}
Eq. (\ref{eqn:chim}) is the {\it mechanical susceptibility} of the moving element. The mechanical linear response is thus modified by the interaction with the microwave field through the term $\Sigma$ [Eq. (\ref{eqn:selfE})]; it is usually (abusively) referred to as the {\it optical ``self-energy''}, see e.g. Ref. \citep{aspelmeyer2014cavity}. Matching the expressions of this Review, 
note the difference in the definition of susceptibilities between mechanical and optical fields [ Eqs. (\ref{eqn:chiopt1}-\ref{eqn:chiopt3})]: an $i$ factor has been incorporated in between.
The last Eq. (\ref{eqn:noise}) corresponds to the {\it stochastic component} of the back-action: noise originating from the Johnson-Nyquist current that adds up with the Langevin force.

Taking real and imaginary parts of $\Sigma$, we see from Eq. (\ref{eqn:chim}) that the optomechanical interaction is responsible for a frequency shift $\delta \Omega_m$ and an additional damping term $\Gamma_{\text{opt}}$:
\begin{eqnarray}
\delta \Omega_m     & = & G^2\frac{1}{\omega_c \, (2 m \Omega_m)} \left( \frac{C_t \omega_c^2 \left|\mu_p \right|^2 }{2} \right) \times \nonumber \\ 
&& \!\!\!\!\!\!\!\!\!\!\!\!\!\!\!\!\!\! \left[ \frac{\Delta+ \Omega_m}{(\Delta+ \Omega_m)^2+(\frac{\kappa_t}{2})^2}+\frac{\Delta- \Omega_m}{(\Delta- \Omega_m)^2+(\frac{\kappa_t}{2})^2} \right] \label{eqn:shift} \! , \\
\Gamma_{\text{opt}} & = & G^2\frac{1}{\omega_c\, (2 m \Omega_m)} \left( \frac{C_t \omega_c^2 \left|\mu_p \right|^2 }{2} \right) \times \nonumber \\ 
&& \!\!\!\!\!\!\!\!\!\!\!\!\!\!\!\!\!\! \left[ \frac{\kappa_t}{(\Delta+ \Omega_m)^2+(\frac{\kappa_t}{2})^2}-\frac{\kappa_t}{(\Delta- \Omega_m)^2+(\frac{\kappa_t}{2})^2} \right] \label{eqn:damp} \! .
\end{eqnarray}
The former expression above is referred to as the {\it optical spring} and the latter the {\it optical damping} effects \citep{aspelmeyer2014cavity}.
Physically, these effects originate in the {\it radiation pressure} exerted on the movable capacitor by the electromagnetic field confined inside it.

In the following Section we shall discuss the spectra associated to Eq. (\ref{eqn:linear}) with their specific properties. The link between injected power, Brownian motion and measured spectrum of the voltage $V_{\text{out}}$ is finally presented.
\subsection{\label{sec:output} Spectral properties and input-output relationships}

The spectrum of the stochastic back-action term [Eq. (\ref{eqn:noise})] writes:
\begin{eqnarray}
\label{eqn:eqn:spectral_force}
S_{\delta F_0} (\omega)& = & \frac{G^2}{\omega_c^2} \left( \frac{C_t \omega_c^2 \left|\mu_p \right|^2 }{2} \right)  \frac{R_t S_{\delta I_n}}{2} \times \nonumber \\   
& &\!\!\!\!\!\!\!\!\!\!\!\!\!\!\!\!\!\!\!\!\!\!\!\!\!\!\! \left[ \frac{\kappa_t}{(\Delta+ \Omega_m)^2+(\frac{\kappa_t}{2})^2}+\frac{\kappa_t}{(\Delta- \Omega_m)^2+(\frac{\kappa_t}{2})^2}  \right] \! , \label{baforce}
\end{eqnarray} 

as a function of the white current noise spectrum $S_{\delta I_n}=S_{\delta I_l}(\omega)=S_{\delta I_h}(\omega)$, with 'l' and 'h' components uncorrelated; therefore $S_{\delta F_0}$ is also white.
As a result, the displacement spectrum deduced from Eq. (\ref{eqn:linear}) is:
\begin{equation}
\label{eqn:mainspectra}
S_{x_0}(\omega) = \left|\chi_m (\omega) \right|^2 \left[ S_{L_0} + S_{\delta F_0} \right] ,
\end{equation}
with $S_{L_0}$ the white force spectrum associated to Brownian motion.
This result is a {\it Lorentzian peak} (depicted in Fig. \ref{fig:schematicFreq} on the left) with full-width $\Gamma_m+\Gamma_{\text{opt}}$ and position $\delta \Omega_m$ (in the frame rotating at $\Omega_m$); the total area is proportional to the total white noise force felt by the mechanics, namely $S_{{\cal L}_0}=S_{L_0} + S_{\delta F_0}$. 
Equivalently, Eq. (\ref{eqn:mainspectra}) writes with the original spectra in the laboratory frame:
\begin{equation}
 S_{x}(\omega) = \left[ \left| \chi_m (\omega-\Omega_m) \right|^2 + \left|\chi_m (\omega+\Omega_m) \right|^2\right]  S_{\cal L} ,
\end{equation}
defined for $\omega$ ranging from $-\infty$ to $+\infty$; the classical spectrum is even with two identical peaks $S^{-}_x(\omega)$, $S^{+}_x(\omega)$ located at $\omega \approx \pm \Omega_m$. 
From the rotating wave transform, we have $S_{L_0}/4=S_{L}=2 k_B T_m \, m \Gamma_m$ and $S_{\delta I_n}/4=S_{I_{\text{noise}}}=2 k_B T_c/R_t$ (with $k_B$ Boltzmann's constant) \citep{clerk2010introduction,aspelmeyer2014cavity}; 
by construction, each fluctuating current $\delta I_n (t)$ is defined over a bandwidth of order $\Omega_m$ (while $I_{\text{noise}}$ covers ${\rm I\!R}$). 
\begin{figure}
\includegraphics[width=0.47\textwidth]{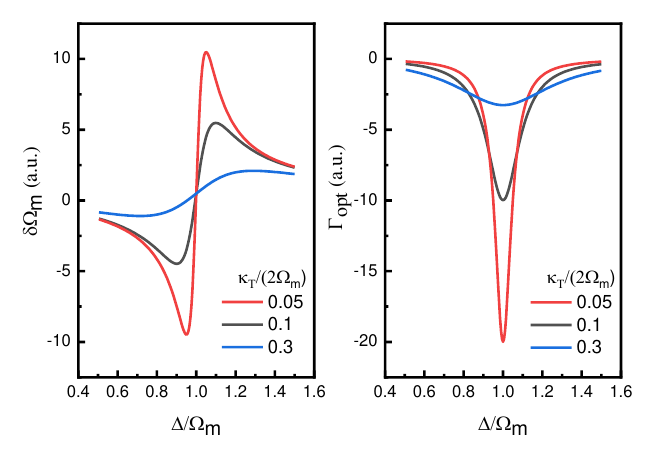}%
\caption{\label{fig:amplif} Illustration (arbitrary units) of the dynamical back-action properties, in the ``blue sideband'' pumping case, for different $\kappa_t/2\Omega_m$ ratios (resolved sideband, see legend). Left: optical spring $\delta \Omega_m$ versus normalized detuning $\Delta/\Omega_m$. Right: optical damping $\Gamma_{\text{opt}}$. The amplitudes of both effects scale as $g^2 \propto \left|\mu_p \right|^2$: When the total damping $\Gamma_m+\Gamma_{\text{opt}}$ reaches $0$, the system starts to self-oscillate (see text).}
\end{figure}
The stochastic force acting on the mechanics can thus be recasted in the simple form:
\begin{equation}
\label{eqn:totnoise}
  \frac{S_{{\cal L}_0}}{4}=S_{{\cal L}}= 2 k_B \,m \left(\Gamma_m+\Gamma_{\text{opt}} \right) T_{\text{eff}} ,
\end{equation}
with: 
\begin{eqnarray}
T_{\text{eff}} & = & \frac{T_m \, \Gamma_m + T_c \, \Gamma_{\text{opt}}'}{\Gamma_m+\Gamma_{\text{opt}}} , \\
\Gamma_{\text{opt}}' & = & G^2\frac{1}{\omega_c\, (2 m \Omega_m)} \left( \frac{C_t \omega_c^2 \left|\mu_p \right|^2 }{2} \right) \frac{\Omega_m}{\omega_c} \times \nonumber \\ 
&& \!\!\!\!\!\!\!\!\!\!\!\!\!\!\!\!\!\! \left[ \frac{\kappa_t}{(\Delta+ \Omega_m)^2+(\frac{\kappa_t}{2})^2}+\frac{\kappa_t}{(\Delta- \Omega_m)^2+(\frac{\kappa_t}{2})^2} \right] \label{eqn:badamp} \! .
\end{eqnarray}
The term $T_{\text{eff}}$ in Eq. (\ref{eqn:totnoise}) is thus interpreted as an {\it effective temperature} for the mechanical mode, created by the combination of the force fluctuations around $\Omega_m$ (in the radiofrequency domain) and the current fluctuations around $\omega_c$ (in the microwave domain), both derived within the same physical framework: the {\it fluctuation-dissipation theorem}. Note the similarity between Eq. (\ref{eqn:damp}) and Eq. (\ref{eqn:badamp}). Besides, we see that the magnitudes of optical spring, optical damping, and back-action noise are all governed by a single parameter:

\begin{equation}
\label{eqn:g}
 g = G \sqrt{\frac{1}{\omega_c\, (2 m \Omega_m)} \left( \frac{C_t \omega_c^2 \left|\mu_p \right|^2 }{2} \right)} .
\end{equation}
Replacing $g$ in the above expressions, they are formally equivalent to the optomechanics results \citep{aspelmeyer2014cavity}; this fact shall be discussed in the following Section.\\

Depending on the scheme used, Eqs. (\ref{eqn:shift},\ref{eqn:damp},\ref{eqn:badamp}) behave very differently. 
When $\Delta \approx + \Omega_m$, the second term in the brackets of these expressions dominate and $\delta \Omega_m \approx g^2 \, (\Delta-\Omega_m)/(\kappa_t/2)^2$, $\Gamma_{\text{opt}} \approx - g^2 \, 4/\kappa_t$ and $\Gamma_{\text{opt}}' \approx g^2 \, \frac{\Omega_m}{\omega_c} 4/\kappa_t$ (resolved sideband limit).
The optical damping is negative 
(the Lorentz mechanical peak has a smaller {\it effective width} $\Gamma_{\text{eff}} = \Gamma_m - |\Gamma_{\text{opt}}|$), 
therefore the mechanical response is enhanced (the mechanical $Q$ factor grows), and the effective temperature $T_{\text{eff}}$ is increased: energy is pumped into the mechanical mode, a mechanism called {\it Stokes scattering} in optics \citep{aspelmeyer2010quantum}. We adopt the language used in this community and call this scheme the ``blue sideband'' pumping. When $\Gamma_m+\Gamma_{\text{opt}} =0$ the system reaches an instability and starts to self-oscillate \citep{aspelmeyer2014cavity,marquardt2006dynamical,carmon2005temporal}. 
The properties of this scheme are illustrated in Fig. \ref{fig:amplif}. When $\Delta \approx - \Omega_m$, the situation is reversed and $\delta \Omega_m \approx g^2 \, (\Delta+\Omega_m)/(\kappa_t/2)^2$, $\Gamma_{\text{opt}} \approx + g^2 \, 4/\kappa_t$ and $\Gamma_{\text{opt}}' \approx g^2 \, \frac{\Omega_m}{\omega_c} 4/\kappa_t$. The optical damping is now positive
($\Gamma_{\text{eff}} = \Gamma_m + |\Gamma_{\text{opt}}|$ is larger), 
the mechanical response is damped (the mechanical $Q$ factor decreases) and the temperature $T_{\text{eff}}$ is reduced: energy is pumped out of the mechanical mode, a mechanism called {\it Anti-Stokes scattering} in optics \citep{aspelmeyer2010quantum}. This scheme, known as {\it sideband cooling} is also referred to as ``red sideband'' pumping. 
At last, when $\Delta \approx 0$ the situation is symmetric: no energy is pumped in or out, and 
$\delta \Omega_m \approx g^2 \, (2 \Delta)/(\Omega_m)^2$, $\Gamma_{\text{opt}} \approx 0$ and $\Gamma_{\text{opt}}' \approx g^2 \, \frac{\Omega_m}{\omega_c} 2 \kappa_t/(\Omega_m)^2$
($\Gamma_{\text{eff}} = \Gamma_m $ unchanged). 
In order to distinguish it from the two other schemes, we name it ``green sideband'' pumping. Note that this scheme has the smallest back-action contribution; it is thus also referred to as the {\it optimal scheme} in optics \citep{aspelmeyer2014cavity}. \\

The final step of the modeling requires to link the input (the $V_{\text{rf}}$ source) to the measured spectrum of the output voltage $S_{V_{\text{out}}}(\omega)$. From Eq. (\ref{eqn:element3}), 
and including the voltage noise on the detector $V_{\text{noise}}(t)$, we have:
\begin{eqnarray}
\label{eqn:vout}
S_{V_{\text{out}}}(\omega) & = & S_{V_{\text{noise}}} + 
\omega_c^4 (C_c Z_0)^2 \times \nonumber \\
&& \!\!\!\!\!\!\!\!\!\!\!\! \sum_{n=-\infty}^{+\infty} \left[\frac{S_{\mu_n}(\omega-\omega_n)}{4} + \frac{S_{\mu_n}(\omega+\omega_n)}{4} \right] \nonumber \\
& & - \omega_c^2 (C_c Z_0) \times \nonumber \\
& & \!\!\!\!\!\!\!\!\!\!\!\!\!\!\!\!\!\!\!\!\!\!\!\!\!\!\!\!\!\!\!\!\!\!\!\!    \sum_{n=-\infty}^{+\infty} \left[ \frac{S_{\mu_n,\delta V_n}(\omega-\omega_n)}{4} + \frac{S_{\mu_n,\delta V_n}(\omega+\omega_n)}{4}  \right] \! ,
\end{eqnarray}
reminding $\omega_n = \omega_p+ n \Omega_m$, with $S_{\mu_n}(\omega)$ the spectrum of the $n^{\text{th}}$ component of the flux $\phi(t)$ decomposition, Eq. (\ref{eqn:ansatz}), and $S_{\mu_n,\delta V_n}(\omega)$ the {\it cross-correlations} between flux and voltage noise components (with $V_{\text{noise}}(t)=\sum_n \frac{1}{2} \delta V_n (t) e^{-i \left(\omega_p+ n \Omega_m\right) t}+ \text{c.c.} $, same decomposition as for the current $I_{\text{noise}}$).
By construction, from the admittance $Y_c$ introduced in Section \ref{sec:soluce}, we have:
\begin{equation}
V_{\text{noise}} \approx + i \omega_c C_c Z_0 R_{ex} \, I_{\text{noise}}  , 
\end{equation} 
which defines the output voltage noise from the cavity current noise.
We have:
\begin{eqnarray}
S_{\mu_p,\delta V_p}(\omega) & =  & + \frac{2(\omega_c C_c Z_0)}{\omega_c C_t} \left[\chi_p + \chi_p^* \right] \frac{R_{ex} S_{\delta I_n} }{4} , \\
S_{\mu_l,\delta V_l}(\omega) & =  & + \frac{2(\omega_c C_c Z_0)}{\omega_c C_t} \left[\chi_l + \chi_l^* \right] \frac{R_{ex} S_{\delta I_n} }{4} \label{bckgdCorr1} \\ \nonumber 
&\!\!\!\!\!\!\!\!\!\!\!\!\!\!\!\!\!\!\!\!\!\!\!\!\! +& \!\!\!\!\!\!\!\!\!\!\!\!\!\!\! \left|\mu_p \right|^2 G^2 (\omega_c C_c Z_0)\left[i(\chi_l^2 \chi_m^*-{\chi_l^*}^2 \chi_m) \right] \frac{R_{ex} S_{\delta I_n} }{4}  ,  \\ 
S_{\mu_h,\delta V_h}(\omega) & =  & + \frac{2(\omega_c C_c Z_0)}{\omega_c C_t} \left[\chi_h + \chi_h^* \right] \frac{R_{ex} S_{\delta I_n} }{4} \label{bckgdCorr2} \\ \nonumber 
&\!\!\!\!\!\!\!\!\!\!\!\!\!\!\!\!\!\!\!\!\!\!\!\!\! +& \!\!\!\!\!\!\!\!\!\!\!\!\!\!\!  \left|\mu_p \right|^2 G^2 (\omega_c C_c Z_0)\left[i(\chi_h^2 \chi_m-{\chi_h^*}^2 \chi_m^*) \right] \frac{R_{ex} S_{\delta I_n} }{4}  , 
\end{eqnarray}
using the properties of the Johnson-Nyquist current. On the right-hand-side, the first term involves only the microwave cavity; for Eqs. (\ref{bckgdCorr1},\ref{bckgdCorr2}) the last term involves the mechanics. These terms are nonzero since {\it the same current noise} generating the detection background also drives the cavity, and is fed back to the mechanics from Eq. (\ref{eqn:noise}). 

The total spectrum $S_{V_{\text{out}}}(\omega)$ is composed of identical combs around $\omega \approx \pm \omega_p$.
What is measured by any classical apparatus (say, a spectrum analyzer) is the {\it power spectral density} [in Watt/(radian/second)]:
\begin{eqnarray}
S_{\text{PSD}}(\omega) & = & \frac{2 S_{V_{\text{out}}}(\omega > 0) }{Z_0} = 2 R_{ex} S_{I_{\text{noise}}} \nonumber \\ \label{eqn:PSD}
&& \!\!\!\!\!\!\!\!\!\!\!\!\!\!\!\!\!\!\!\!\!\! + \frac{\omega_c^4 (C_c Z_0)^2}{2 Z_0} \times \sum_{n='p','l','h'} S_{\mu_n}(\omega-\omega_n) \nonumber \\
&& \!\!\!\!\!\!\!\!\!\!\!\!\!\!\!\!\!\!\!\!\!\! - \frac{\omega_c^2 (C_c Z_0)}{2 Z_0} \times \sum_{n='p','l','h'} S_{\mu_n,\delta V_n}(\omega-\omega_n)    ,
\end{eqnarray}
with all power folded in the $\omega > 0$ range, since the classical noise spectral density is symmetric in frequency \citep{clerk2010introduction}. 
Similarly to the cavity itself, we define a temperature for the detection port as $R_{ex} S_{I_{\text{noise}}}=R_{ex} S_{\delta I_n}/4=k_B T_{ex}$, ensuring that the background noise in Eq. (\ref{eqn:PSD}) reduces to $2 k_B T_{ex}$, as it should.  
In the sum of Eq. (\ref{eqn:vout}), only the 'p', 'l' and 'h' terms have been kept: the measured spectrum is composed of 3 peaks (see Figs. \ref{fig:schematicFreq} and \ref{fig:sideband}), defined from Eqs. (\ref{eqn:ampInside}):
\begin{eqnarray}
\!\!\!\!\!\!\!\!\!\!\!\! S_{\mu_p}(\omega) &\!\! = &\! \left|\mu_p \right|^2 \, 2 \pi \delta_0(\omega) + \frac{R_t S_{\delta I_n} }{4} \frac{\kappa_t}{C_t \omega_c^2} \left| \chi_p \right|^2 \!\! , \label{eqn:pumppeak} 
\end{eqnarray}
\begin{eqnarray}
\!\!\!\!\!\!\!\!\!\!\!\! S_{\mu_l}(\omega) &\!\! = &\! \left|\mu_p \right|^2 G^2 \frac{S_{x_0}(\omega)}{4} \left| \chi_l \right|^2 + \frac{R_t S_{\delta I_n} }{4} \frac{\kappa_t}{C_t \omega_c^2} \left| \chi_l \right|^2 \nonumber \\
& & \!\!\!\!\! + \frac{ \left|\mu_p \right|^2 G^2 \kappa_t }{2 \omega_c} \left[ i( \chi_m^* \chi_l - \chi_m \chi_l^* )\right] \frac{R_t S_{\delta I_n} }{4} \left| \chi_l \right|^2 \! , \label{eqn:lpeak} \\
\!\!\!\!\!\!\!\!\!\!\!\! S_{\mu_h}(\omega) &\!\! = &\!\! \left|\mu_p \right|^2 G^2 \frac{S_{x_0}(\omega)}{4} \left| \chi_h \right|^2  + \frac{R_t S_{\delta I_n} }{4} \frac{\kappa_t}{C_t \omega_c^2} \left| \chi_h \right|^2 \nonumber \\
& & \!\!\!\!\!\!\!\!\!\! + \frac{ \left|\mu_p \right|^2 G^2 \kappa_t }{2 \omega_c} \left[ i( \chi_m \chi_h - \chi_m^* \chi_h^* )\right] \frac{R_t S_{\delta I_n} }{4} \left| \chi_h \right|^2 \!  , \label{eqn:hpeak}
\end{eqnarray}
applying again the properties of the Johnson-Nyquist current ($\delta_0$ is the Dirac function).
The second terms in each expressions correspond to the cavity alone, being driven by the current noise. 
Eq. (\ref{eqn:pumppeak}) is due to the pump tone signal; Eqs. (\ref{eqn:lpeak},\ref{eqn:hpeak}) include the two sidebands, proportional to the mechanical motion spectrum and $\left|\mu_p \right|^2 = \frac{\left| I_p \right|^2}{4 (\omega_c C_t)^2} \left| \chi_p \right|^2$.
The last terms in  Eqs. (\ref{eqn:lpeak},\ref{eqn:hpeak}) correspond to cross-correlations between the cavity noise current and the motion. \\

We should now clarify the energy flow in this system. The power injected $P_{in}$ by the traveling wave $\phi_{in}$ (Fig. \ref{fig:elementCircuit}) is by definition:
\begin{equation}
P_{in} = \frac{1}{2} \frac{\left| V_p \right|^2}{Z_0} ,
\end{equation}
and the energy $E_c$ stored in the microwave resonator writes:
\begin{equation}
\label{eqn:ec}
E_c = \frac{C_t \omega_c^2 \left|\mu_p \right|^2 }{2} = P_{in} \, \kappa_{ex} \left|\chi_p \right|^2,
\end{equation}
where we made use of Eqs. (\ref{eqn:element1},\ref{eqn:element2},\ref{eqn:kappaex}). The pump power $P_{pump}$ measured in the output spectrum at $\omega_p$ is then: 
\begin{equation}
\label{eqn:ppump}
P_{pump} =  \frac{\omega_c^4 (C_c Z_0)^2}{2 Z_0} \left|\mu_p \right|^2 = E_c  \, \kappa_{ex} .
\end{equation}
Replacing Eq. (\ref{eqn:ec}) in Eq. (\ref{eqn:ppump}) leads to $P_{pump} =P_{in} \, \kappa_{ex}^2 \left|\chi_p \right|^2$; the ratio $P_{pump} / P_{in}$ is thus a straightforward calibration of the quantity $\kappa_{ex}$.  %
%
The power spectral density can thus be recasted in the compact form:
\begin{eqnarray}
\label{eqn:psdfinal}
&& \!\!\!\!\!\!\!  S_{\text{PSD}}(\omega) = 2 k_B T_{ex} +  P_{pump} \, 2 \pi \delta_0(\omega-\omega_p) \nonumber \\
&&\,\,\,\,\,\,\,\,\,\,\,\,\,\,\,\,\,\, + \kappa_{ex} \, k_B (T_c-T_{ex}) \, \kappa_t \left| \chi_p \right|^2 \Pi (\omega-\omega_p) \nonumber \\
&&+  \kappa_{ex} \, g^2 \, \frac{S^{-}_{x}(\omega-\omega_p)}{\bar{x}^2} \left| \chi_l \right|^2  \nonumber \\
&&\,\,\,\,\,\,\,\,\,\,\,\,\,\,\,\,\,\, + \kappa_{ex} \, k_B (T_c-T_{ex}) \, \kappa_t \left| \chi_l \right|^2 \Pi (\omega-\omega_l)\nonumber \\
&&+  \kappa_{ex} \, g^2 \, \frac{S^{+}_{x}(\omega-\omega_p)}{\bar{x}^2} \left| \chi_h \right|^2 \nonumber \\
&&\,\,\,\,\,\,\,\,\,\,\,\,\,\,\,\,\,\, + \kappa_{ex} \, k_B (T_c-T_{ex}) \, \kappa_t \left| \chi_h \right|^2 \Pi (\omega-\omega_h), \label{finalPSD}
\end{eqnarray}
having defined $\bar{x}^2=\frac{1}{\omega_c\, (2 m \Omega_m)}$ (in meter$^2$/Joule). $\Pi(\omega)$ denotes the door function (here, of width $\sim \Omega_m$), reminding that each cavity component is defined around a precise angular frequency $\omega_n$.
Note that to detect the cavity as a peak or a dip, one requires $T_c \neq T_{ex}$; 
$T_{ex}$ also defines the {\it background noise level} that ultimately limits a measurement. 
Eq. (\ref{eqn:g}) then reads $g= G \,\bar{x} \sqrt{E_c}$ (in radian/second).
%

The last terms of
Eqs. (\ref{bckgdCorr1},\ref{eqn:lpeak}) and (\ref{bckgdCorr2},\ref{eqn:hpeak}), which correspond to the cross-correlations involving the mechanics, 
affect the measurement by {\it mimicking} an extra force noise $\delta F_{ex\,l}, \delta F_{ex\,h}$ which depends on the sideband \citep{aspelmeyer2014cavity,weinstein2014observation}:
\begin{eqnarray}
S^{-}_{x}(\omega) & = & S_{x}(\omega) + |\chi_m (\omega - \Omega_m)|^2 S_{\delta F_{ex\,l} }, \label{SSFl}   \\ 
S^{+}_{x}(\omega) & = & S_{x}(\omega) + |\chi_m (\omega + \Omega_m)|^2 S_{\delta F_{ex\,h} }. \label{SSFh}
\end{eqnarray}
These shall not be confused with the true back-action force noise $S_{\delta F}$: $\delta F_{ex\,l}, \delta F_{ex\,h}$ do not actually affect the mechanical degree of freedom. Their relevance is discussed in Section \ref{sec:asymmetry}, on the basis of the 3 standard measuring schemes (with i.e. $\Delta= 0, \pm \Omega_m$).
Notwithstanding this fact, the two sidebands are thus the image of the two peaks of the mechanical spectrum, translated around $\omega_p$ (one being thus at $\omega_p-\Omega_m = \omega_l$ and the other at $\omega_p+\Omega_m = \omega_h$), with an amplitude proportional to $g^2$ and modulated by the susceptibilities $\left| \chi_l \right|^2, \left| \chi_h \right|^2$. %
 Integrating the peaks, we obtain an area proportional to the observed variance of the displacement $\frac{1}{2 \pi} \int S^{\pm}_{x} d\omega = {\sigma^{\pm}_x}^2 $, including thus the cross-correlation contribution. \\

The above applies to the reflection setup, Fig. \ref{fig:lcr_parelle}  (c). For the two-port one Fig. \ref{fig:lcr_parelle}  (a), 
one should replace $\kappa_{ex} \rightarrow \kappa_1$ in Eq. (\ref{eqn:ec}) and $\kappa_{ex} \rightarrow \kappa_2$ in Eqs. (\ref{eqn:ppump},\ref{eqn:psdfinal}). In the case of a bi-directional arrangement Fig. \ref{fig:lcr_parelle}  (b),  one should replace $\kappa_{ex} \rightarrow \kappa_{ex}/2$ in all expressions.
Up to this point, we relied only on {\it classical mechanics}, and all optomechanical properties (at fixed $\Delta$) depend only on $g$ (tuned experimentally through $P_{in}$) and $T_m, T_c, T_{ex}$. We shall now explicitly link our results to the quantum formalism.
\section{\label{sec:discussion} Discussion}

\subsection{\label{sec:asymmetry} Sideband asymmetry}

Cross-correlations between the cavity current noise and the mechanics, Eqs. (\ref{eqn:lpeak},\ref{eqn:hpeak}) and cross-correlations between the detection background and the cavity noise Eqs. (\ref{bckgdCorr1},\ref{bckgdCorr2}) can be recast into {\it apparent stochastic force components} that depend on the sideband, Eqs. (\ref{SSFl},\ref{SSFh}) for the 'l' and 'h' ones respectively. 

For the ``blue'' pumping scheme, only the 'l' sideband is measurable in the sideband-resolved limit. Injecting $\Delta = + \Omega_m$ in the above mentioned equations, we obtain:
\begin{equation}
S_{\delta F_{ex\,l} } = 2 m \Gamma_{\text{eff}} \, k_B \left(+2 T_c - T_{ex} \right)\frac{\Omega_m}{\omega_c} ,\label{SAblue}
\end{equation}
with $T_c$ and $T_{ex}$ the temperatures of the cavity and the detection port respectively, as introduced in the preceding Section.
Similarly for the ``red'' scheme, with $\Delta = - \Omega_m$ and looking at the 'h' sideband we have:
\begin{equation}
S_{\delta F_{ex\,h} } = 2 m \Gamma_{\text{eff}} \, k_B \left(+T_{ex} - 2 T_c \right) \frac{\Omega_m}{\omega_c}.\label{SAred}
\end{equation}
In both expressions, $\Gamma_{\text{eff}}=\Gamma_m+\Gamma_{\text{opt}}$ but $\Gamma_{\text{opt}}$ is different: negative for the ``blue'' scheme, and positive for the ``red'' one. However, for low drive powers $\Gamma_{\text{opt}} \approx 0$ and $\Gamma_{\text{eff}} \approx \Gamma_m$. In this case, a very simple result emerges: the two apparent force noises are opposite, a result referred to in the literature as {\it sideband asymmetry} \citep{aspelmeyer2014cavity,weinstein2014observation}.
But here, we remind the reader that the feature is {\it purely classical}, and by no means a signature of quantum fluctuations.

In the case of a ``green'' pumping scheme,  $\Delta = 0$ and both sidebands can be measured at the same time.
The resulting expressions for the cross-correlation apparent stochastic force components are:
\begin{eqnarray}
S_{\delta F_{ex\,l} } &=& 2 m \Gamma_{m} \, k_B \left(+ T_{ex} \right) \frac{\Omega_m}{\omega_c}, \nonumber \\
S_{\delta F_{ex\,h} } &=& 2 m \Gamma_{m} \, k_B \left(- T_{ex} \right) \frac{\Omega_m}{\omega_c} , \label{SAgreen}
\end{eqnarray}
again in the sideband-resolved limit. %
Eqs. (\ref{SAgreen}) are very similar to Eqs. (\ref{SAblue},\ref{SAred}): again the two forces are opposite, but this time they depend only on $T_{ex}$. \\
%

Let us consider the case of an ideally thermalized system were $T_c=T_{ex}=T_m$. 
Then in the limit $\Gamma_{\text{opt}} \approx 0$, sideband asymmetry measured by comparing the 'l' peak in ``blue'' pumping Eq. (\ref{SAblue}) with the 'h' peak in ``red'' Eq. (\ref{SAred}) gives strictly the same result as the direct comparison of the two sidebands Eqs. (\ref{SAgreen}) observed with a ``green'' scheme.
Besides, the sideband asymmetry effect simply renormalizes the observed mechanical temperature by $T_m \rightarrow T_m (1+\Omega_m/\omega_c)$ on the 'l' side, and by $T_m \rightarrow T_m (1-\Omega_m/\omega_c)$ on the 'h' side; since $\Omega_m/\omega_c \ll 1$, this effect can be safely neglected in this case. 
One needs to artificially create a situation where $T_m \ll T_{ex}$ to make the sideband asymmetry detectable (e.g. by sideband cooling the mechanical mode, and injecting noise through the microwave port) \citep{weinstein2014observation}.
As soon as $T \rightarrow 0~$K, the classical picture breaks down and all features should be interpreted in the framework of quantum mechanics; including sideband asymmetry. The direct link between the two theories shall be discussed in the next Section.

\subsection{\label{sec:Heisenberg} The quantum limit and Heisenberg uncertainty}

Quantum mechanics tells us that energy comes in {\it quanta}, each carrying $\hbar$ times the frequency corresponding to the concerned mode. 
To link the classical writing to the quantum expressions, we thus have to introduce the following {\it populations}:
\begin{eqnarray}
n_c & = & \frac{E_c}{\hbar \omega_c} , \\
n_c^{th} & = & \frac{k_B T_c}{\hbar \omega_c} , 
\end{eqnarray}
\begin{eqnarray}
n_{ex}^{th} & = & \frac{k_B T_{ex}}{\hbar \omega_c} , \\
n_m^{th} & = & \frac{k_B T_m}{\hbar \Omega_m} ,
\end{eqnarray}
with respectively the (coherent) cavity population, the cavity thermal population, the external microwave port (thermal) population and the mechanical mode thermal population. 

Let us consider first the case of a mechanical oscillator driven only by the Langevin force; we do not consider yet the back-action stochastic drive, neither the sideband asymmetry apparent contributions discussed in Section \ref{sec:asymmetry}.
Then 
$\frac{1}{2 \pi} \int S^{\pm}_{x} d\omega = {\sigma^{\pm}_x}^2 = {\sigma_x}^2 $, the two detected sidebands are equivalent, and we have:
\begin{eqnarray}
{\sigma_x}^2 & = & \left\langle {\delta x}^2 \right\rangle \frac{\Gamma_m}{\Gamma_{\text{eff}}} , \label{barecase} \\
 \left\langle {\delta x}^2 \right\rangle & = & \frac{k_B T_m} {2 m {\Omega_m}^2}  = x_{zpf}^2 \,n_m^{th} , \label{x2}
\end{eqnarray}
with $\left\langle {\delta x}^2 \right\rangle$ the half-variance of the motion (computed on one sideband only), having defined $x_{zpf} = \sqrt{\hbar/(2 m \Omega_m)}$ the {\it zero point fluctuations}.
The $\Gamma_m/\Gamma_{\text{eff}}$ factor comes from the dynamical part of the back-action, causing optical damping/anti-damping, with $\Gamma_{\text{eff}}=\Gamma_m+\Gamma_{\text{opt}}$ for ``blue'' and ``red'' pumping schemes, and $\Gamma_{\text{eff}}=\Gamma_m$ for ``green''.
From this definition we have $x_{zpf}^2 = \hbar \omega_c \, \bar{x}^2$, and we can recast $g =g_0 \sqrt{n_c}$ with $g_0= G \, x_{zpf}$ (in Rad/s).
Eq. (\ref{x2}) coincides exactly with the quantum-mechanical high-temperature limit.  
But when $T \rightarrow 0~$K, $n_m^{th}\rightarrow 0$ and 
$n_m^{th}$ should be replaced by the factor $1/2$ which corresponds to the vacuum noise predicted by quantum mechanics \citep{clerk2010introduction,aspelmeyer2014cavity}. In the following, the same treatment shall be performed for $n_c^{th}$ in the so-called quantum ($T \rightarrow 0~$K) limit.  

Any measurement comes with an acquisition imprecision. 
In the literature, one finds a discussion focused on the phase of the optical ``green'' readout \citep{aspelmeyer2014cavity}.
An equivalent discussion can be performed on the {\it amplitude} of the signal; this is what we will present in the following.
A position fluctuation $\delta x$ transduces into a cavity frequency shift $\delta \omega_c =\delta x\, G\, (\kappa_t/2) /{\left|-i \omega+(\kappa_t/2)\right|}$, taking into account the finite response time of the microwave mode. This shift $\delta \omega_c $ will in turn modify the output signal energy $\delta E/E \approx  4 {\left(\delta \omega_c \right)}^2/{\kappa_t}^2 $. 
This variation can be expressed in terms of {\it detection noise} quanta $\delta E/ (\hbar \omega_c) = n_{det}$ with 
$\delta E/E =n_{det}/\left( n_c \kappa_c \,t \right)$, measured during a time $t$. For an ideal quantum detector the measurement is shot-noise limited with $n_{det}=1$ \citep{aspelmeyer2014cavity,clerk2010introduction}; in the classical case, $n_{det} \, \hbar \omega_c$ corresponds to the noise background (in Joules) affecting the detection, arising from the whole amplification chain (with obviously $n_{det}\gg 1$).
The imprecision in position resulting from the finite $n_{det}$ can be interpreted as the integral of a flat noise $S^{imp}_{x}$ (over a bandwidth $t^{-1}$) \citep{Schliesser2009, aspelmeyer2014cavity}:
\begin{equation}
S^{imp}_{x} = \frac{k^2_t n_{det}}{16 \, G^2 n_c \kappa_{ex}}\left(1+4 \frac{\omega^2}{\kappa_t^2} \right), \label{Simp}
\end{equation}

where $\omega$ is set by the measuring scheme, ``blue'', ``red'' or ``green''.

Focusing on the ``green'' scheme, Eq. (\ref{baforce}) leads to $S_{\delta F}= \hbar^2 G^2 n_c n_c^{th} \,2 \kappa_t/\Omega_m^2$, and in
Eq. (\ref{Simp}) one should consider $\omega = \pm \Omega_m$ (for each sideband). This leads to the result:
\begin{equation}
S^{imp}_{x} \, S_{\delta F} = \frac{\hbar^2}{4} \frac{\kappa_t}{\kappa_{ex}} \left(2 n_c^{th} \right) n_{det}. \label{shit}
\end{equation}
At $T \rightarrow 0~$K, as mentioned above $n_c^{th}$ is replaced by $1/2$. 
Thus for a shot-noise limited detection ($n_{det}=1$), we recover from Eq. (\ref{shit}) the famous {\it Heisenberg limit} $S^{imp}_{x} S_{\delta F}\geq \hbar^2/4$, reached only for an overcoupled cavity $\kappa_t \approx \kappa_{ex}$.

It is enlightening to evaluate the minimal mechanical displacement which can be detected with such a microwave optomechanical scheme. 
The imprecision noise Eq. (\ref{Simp}) can be taken into account by adding it up with $S_x^{\pm}$ in Eq. (\ref{finalPSD}). 
Subtracting the $2 k_B T_{ex}$ background and integrating each sideband over a bandwidth $\Delta \omega$ large enough to cover the peaks, we are led to define 
a signal component ${\cal S}_{ig}$, and a noise component ${\cal N}_{oise}$, for each sideband $i=$'l' or 'h':

\begin{eqnarray}
\label{eqn:SN}
{\cal S}_{ig} & = & \hbar \omega_c |\chi_{i}|^2 \kappa_{ex} \, G^2 n_c \left[ \left\langle {\delta x}^2 \right\rangle \frac{\Gamma_m}{\Gamma_{\text{eff}}}  \right. \nonumber \\
& & \,\,\,\,\,\,\,\,\,\, \left. + \frac{ S_{\delta F_{ex\,i} } }{(2 m \Omega_m)^2 \, \Gamma_{\text{eff}} } \right] , \label{signal} \\
{\cal N}_{oise} & = &  \hbar \omega_c |\chi_{i}|^2  \kappa_{ex} \Bigg[  \Bigg. \nonumber \\
& & \!\!\!\!\! \Bigg. \kappa_t \frac{\Delta \omega}{2 \pi} \left( \frac{n_{det} \, \kappa_t }{16 \, \kappa_{ex}}\left( 1+4 \frac{\omega^2}{\kappa_t^2} \right) + n_c^{th}-n_{ex}^{th} \right) \Bigg.   \nonumber \\
& & \!\!\!\!\!\!\! \Bigg. + \frac{G^4 n_c^2 \, x_{zpf}^4 \kappa_t }{\Gamma_m} n_c^{th} \left(|\chi_l|^2+|\chi_h|^2\right)  \frac{\Gamma_m}{\Gamma_{\text{eff}}} \, \Bigg], \label{noise}
\end{eqnarray}
where ${\cal N}_{oise}$ contains {\it both} imprecision (former term, with also cavity noise $n_c^{th}-n_{ex}^{th}$) and back-action (latter).
We now wrote explicitly the sideband asymmetry contribution $S_{\delta F_{ex\,i} }$ in Eq. (\ref{signal}).

Considering again the ``green'' scheme, 
we have $\left\langle {\delta x}^2 \right\rangle \Gamma_m/\Gamma_{\text{eff}} = x_{zpf}^2 n_m^{th}$ and $|\chi_{l}|^2=|\chi_{h}|^2$.
Eqs. (\ref{SAgreen}) demonstrate that the extra term in Eq. (\ref{signal}) modifies the measured peaks from Eq. (\ref{x2}) into Eq. (\ref{signal}) by substituting $n_m^{th} \rightarrow n_m^{th} + n_{ex}^{th} $ on the 'l' side, and $n_m^{th} \rightarrow n_m^{th} - n_{ex}^{th} $ on the 'h' side; this is sideband asymmetry in the quantum mechanics language.
The {\it difference} ${\sigma^{-}_x}^2-{\sigma^{+}_x}^2$ is then proportional to $ 2 n_{ex}^{th}$, which tends towards 1 in quantum mechanics at $T \rightarrow 0~$K; a similar result can be obtained comparing the 'l' and 'h' peaks obtained in ``blue'' and ``red'' pumping schemes respectively \citep{weinstein2014observation}.  
However from $({\sigma^{-}_x}^2+{\sigma^{+}_x}^2)/2$, the ``green'' scheme leads to a quantity {\it insensitive} to sideband asymmetry, which is directly the image of the mechanical motion.
Discussing now on this quantity, our signal is then ${\cal S}_{ig} \propto \left\langle {\delta x}^2 \right\rangle$.

\begin{figure}
\includegraphics[width=0.45\textwidth]{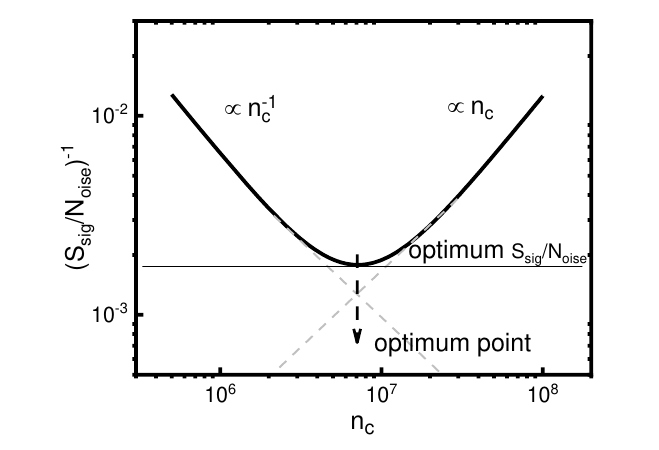}%
\caption{\label{fig:optimal} Illustration of inverse 
 signal-to-noise ratio $\left( {\cal S}_{ig}/{\cal N}_{oise} \right)^{-1}$ in the ``green '' pumping scheme, as a function of photon number $n_c$ inside the cavity. 
Driven at optimal $n_c$, the best signal-to-noise ratio can be obtained for the detection of $\left\langle{{\delta}x}^2\right\rangle$ 
(so-called {\it standard classical limit}, see text). 
Curve computed with 
, $\kappa_t = \kappa_{ex}$,
 $\Delta \omega =6 \, \Gamma_m$, and $\Gamma_m \Omega_m^2/(\kappa_t g_0^2)=1.6\,10^{8}$  %
 for $n_{det}=100$, $n_c^{th} =n_{ex}^{th}=6.\, 10^3$ and $n_m^{th} =6.\, 10^5$.  
At the quantum limit, the optimum point (the minimum of the curve) reaches $\left( {\cal S}_{ig}/{\cal N}_{oise} \right)^{-1} \approx 2$ (SQL, see text).
}
\end{figure}
%

Carrying out the substitutions $\Gamma_{\text{eff}}=\Gamma_m$, $\omega=\pm \Omega_m$ and $|\chi_l|^2+|\chi_h|^2 \approx 2 /\Omega_m^2$ valid for ``green'' pumping in Eqs. (\ref{signal},\ref{noise}), we can define a signal-to-noise ratio ${\cal S}_{ig}/{\cal N}_{oise}$ that illustrates our sensitivity to the quantity $\left\langle {\delta x}^2 \right\rangle$. 
This is represented in Fig. \ref{fig:optimal} as a function of drive power with the parameter $n_c$; similar plots can be found in Refs. \citep{aspelmeyer2014cavity,clerk2010introduction,teufel2009nanomechanical}.
On the left, the sensitivity is lost $\propto 1/n_c$ because of our finite detection noise $n_{det}$. On the right, the measurement is dominated by back-action $\propto n_c$ arising from $n_c^{th}$. There is an {\it optimum} defined by $d ({\cal S}_{ig}/{\cal N}_{oise})/d\, n_c=0$.
This point verifies $n_c \approx \frac{1}{4 \sqrt{\pi}} \sqrt{\frac{\Delta \omega}{\Gamma_m}} \sqrt{\frac{\kappa_t}{\kappa_{ex}}} \sqrt{\frac{n_{det}}{n_c^{th}}} \, \frac{\Gamma_m \Omega_m^2}{\kappa_t g_0^2}$, with  ${\cal S}_{ig}/{\cal N}_{oise} \approx \frac{n_m^{th}}{\sqrt{n_{det} 2n_c^{th}}} \sqrt{2\pi} \sqrt{\frac{\kappa_{ex}}{\kappa_t} } \sqrt{\frac{\Gamma_m}{\Delta \omega}} $ (resolved sideband limit, assuming $n_{ex}^{th} = n_c^{th}$).
At the $T \rightarrow 0~$K quantum limit $n_c^{th}, n_m^{th}$ are replaced by $1/2$; with a shot-noise limited detector $n_{det}=1$, 
we reach at best ${\cal S}_{ig}/{\cal N}_{oise} \approx 1/2$ (for $\kappa_t \approx \kappa_{ex}$, $\Delta \omega \approx 6 \, \Gamma_m$): the signal is about half the total detected noise \citep{clerk2010introduction}. This is called the {\it standard quantum limit}, 
which reaches the ultimate physical limit when simultaneously measuring two non-commuting quadratures of the motion \citep{lei2016quantum}.  
In contrast, the classical optimum %
which we shall call {\it standard classical limit} (SCL) is {\it relative} and depends both on $n_{det}$ (quality of classical detector) and $n_c^{th}$ (Johnson-Nyquist noise of the cavity).
The main optomechanical results applying to the ``green'' pumping scheme are compared in Tab. \ref{tab:table1} in the classical and quantum regimes. 
The key point revealed by the classical modeling is that all features have a classical analogue; only the $T \rightarrow 0~$K {\it quantities} are a true signature of quantumness, which highlights the importance of calibrations in all conducted experiments. \\

\begin{table}
\caption{\label{tab:table1}%
$S^{imp}_{x} \, S_{\delta F}$ product, signal ${\cal S}_{ig}$, noise ${\cal N}_{oise}$ and signal-to-noise ratio in the quantum and classical limits
 (the latter are given at the optimal $n_c$ for the ``green'' pumping scheme; $\delta E$ energy detection resolution and $T_c$ cavity temperature, see text). 
}
\begin{ruledtabular}
\begin{tabular}{lcdr}
\textrm{Quantity}&
\textrm{Quantum limit}&
\textrm{Classical limit}\\
\colrule
$S^{imp}_{x} \, S_{\delta F}$ & $\frac{\kappa_t}{\kappa_{ex}} \hbar^2/4 $ & $$\frac{1}{2} \frac{\kappa_t}{\kappa_{ex}} k_B T_c \, \delta E/(\omega_c^2)$$ \\
${\cal S}_{ig}$ & $\propto x_{zpf}^2/{2}$ & $$\propto k_B T_m/(2 \, m \Omega_m^2)$$ \\
${\cal N}_{oise}$ & $\propto x_{zpf}^2$ & $$\propto \sqrt{\delta E \, k_B T_c}/(\sqrt{2} \, m\Omega_m\omega_c)$$ \\
${\cal S}_{ig}/{\cal N}_{oise}$ & 1/2 & $$k_B  T_m \omega_c/(\sqrt{2} \sqrt{\delta E \, k_B  T_c} \, \Omega_m)$$
\end{tabular}
\end{ruledtabular}
\end{table}

A similar reasoning can be performed for the ``blue'' pumping scheme.
While the measurement is limited by the parametric instability for $n_c$, the ${\cal S}_{ig}/{\cal N}_{oise}$ can be recast into $\sim T_m \omega_c/\left(T_c \Omega_m\right)\gg 1$ close to it. This makes it a very practical technique to perform {\it thermometry} \cite{zhou2019chip}.
%

\section{\label{sec:conclusion}Conclusion}

In summary, we have presented the generic classical electric circuit model which is analogous to the standard optomechanics quantum treatment.  
The developed analytics provides the bridge between circuit parameters and quantum optics quantities, a mandatory link for design and optimization. The two approaches are strictly equivalent, provided temperatures are high enough for both the mechanical and the electromagnetic degrees of freedom. 
We considered here the 3 standard single-tone schemes, for which we present all relevant properties. But the modeling can obviously be extended to more complex schemes (like two-tone drives), and more complex structures (e.g. multi-NEMS/multi-cavities setups).

Besides, a thorough comparison of the two models gives a profound understanding of what measured properties are fundamentally quantum. 
To match the mathematics of the two computation methods, we introduce {\it populations} by means of an energy per quanta proportional to the mode resonance frequency: the early {\it Planck postulate}. 
Sideband asymmetry is derived in classical terms, and we distinguish the temperatures of the mechanical mode $T_m$ from the one of the microwave mode $T_c$ and the microwave (detection) port $T_{ex}$.
Considering the measurement protocol in itself, we derive the resolution limit of a classical experiment performed with the optimal optomechanical scheme.
We obtain the classical (and relative) analogue of the (absolute) {\it standard quantum limit} (SQL) fixed by the {\it Heisenberg principle}
in quantum mechanics; 
we shall name it the {\it standard classical limit} (SCL). 
{\it Only} the $T_m, T_c, T_{ex} \rightarrow 0~$K measured quantities appear to be specific to quantum mechanics, since all features present a classical analogous counterpart.

\begin{acknowledgments}
We would like to acknowledge support from the STaRS-MOC project No. 181386 from Region Hauts-de-France, the ERC generator project No. 201050 from ISITE and the ERC CoG grant ULT-NEMS No. 647917. 
The research leading to these results has received funding from the European Union's Horizon 2020 Research and Innovation Programme, under grant agreement No. 824109, the European Microkelvin Platform (EMP). 

\end{acknowledgments}

\nocite{*}
\bibliography{classic_rev}

\begin{thebibliography}{40}
\bibitem[{not(Author's note)}]{note}
\bibinfo{journal}{Note that impedances are expressed in the standard
  electronics language assuming $A_0 \, \exp(+i \omega t)$ time-dependencies.
  The writing should be adapted for full $A_0 \, \exp(-i \omega t)+
  \text{c.c.}$ expressions.}  (\bibinfo{year}{Author's note}).
\end{thebibliography}

\end{document}